\documentclass[aps,prb,twocolumn,showpacs,byrevtex]{revtex4-1}

 \usepackage[english]{babel}
 \usepackage{amssymb,amsmath}
 \usepackage{color}
 \usepackage{pstricks}
 \usepackage{longtable}
 \usepackage{bm}

\usepackage{bbold}
\usepackage{graphicx}
\usepackage{color}
\usepackage{epsfig}

\begin{document}

\title{Wigner Localization in a Graphene Quantum Dot with a Mass Gap}

\author{K. A. Guerrero-Becerra}
\affiliation{CNR-NANO Research Center S3, Via Campi 213/a, 41125 Modena, 
Italy}
\affiliation{Dipartimento di Scienze Fisiche, Informatiche 
e Matematiche,
Universit\`a degli Studi di Modena and Reggio Emilia, Italy}

\author{Massimo Rontani}
\affiliation{CNR-NANO Research Center S3, Via Campi 213/a, 41125 Modena, Italy}
\email{massimo.rontani@nano.cnr.it}
\homepage{www.nano.cnr.it}

\begin{abstract}
In spite of unscreened Coulomb interactions
close to charge neutrality, relativistic massless electrons in graphene 
allegedly behave as noninteracting particles.
A clue to this paradox is that both interaction and kinetic energies 
scale with particle density in the same way. In contrast, 
in a dilute gas of nonrelativistic electrons  
the different scaling drives the transition to Wigner crystal. 
Here we show that Dirac electrons in a 
graphene quantum dot with a mass gap localize \`a la Wigner
for realistic values of device parameters. Our theoretical evidence relies 
on many-body observables
obtained through the exact diagonalization of the interacting 
Hamiltonian, which allows us to take all electron correlations into account.
We predict that the experimental signatures of Wigner localization are the 
suppression of the fourfold periodicity of the filling sequence
and the quenching of excitation energies, 
which may be both accessed through Coulomb blockade spectroscopy.
Our findings are relevant to other carbon-based nanostructures 
exhibiting a mass gap.
\end{abstract}


\maketitle

\section{Introduction}

The role of electron-electron interactions in graphene is a fundamental
and yet open issue \cite{CastroNeto2009,Abergel2010,DasSarma2011,Kotov2012} 
that impacts on the operation of quantum dots (QDs) 
\cite{Recher2010,Rozhkov2011,Guttinger2012}
and other graphene-based 
nanodevices.\cite{Prezzi2008,Yazyev2010,Rozhkov2011,Wang2011,Ki2012}
Since the density of states vanishes at the charge neutrality point,
making Coulomb interaction unscreened,
one might expect strongly correlated behavior at low energies.
Indeed, the fine structure constant 
$\alpha = e^2/(\epsilon\hbar v_F)$---which is the ratio of Coulomb 
to Fermi energy---is of order unity, 
much larger than the value
$\alpha = 1/137$ of quantum electrodynamics, therefore the many-body problem
may not be treated with perturbative methods
(here $\epsilon$ is the background
dielectric constant and $v_F$ the Fermi velocity). 
As a matter of fact, the 
predicted ratio of viscosity to entropy per electron is characteristic
of an extremely interacting quantum fluid.\cite{Muller2009}

\begin{figure}
\setlength{\unitlength}{1 cm}
\begin{picture}(8.5,7.5)
\put(-0.1,3.3){\epsfig{file= 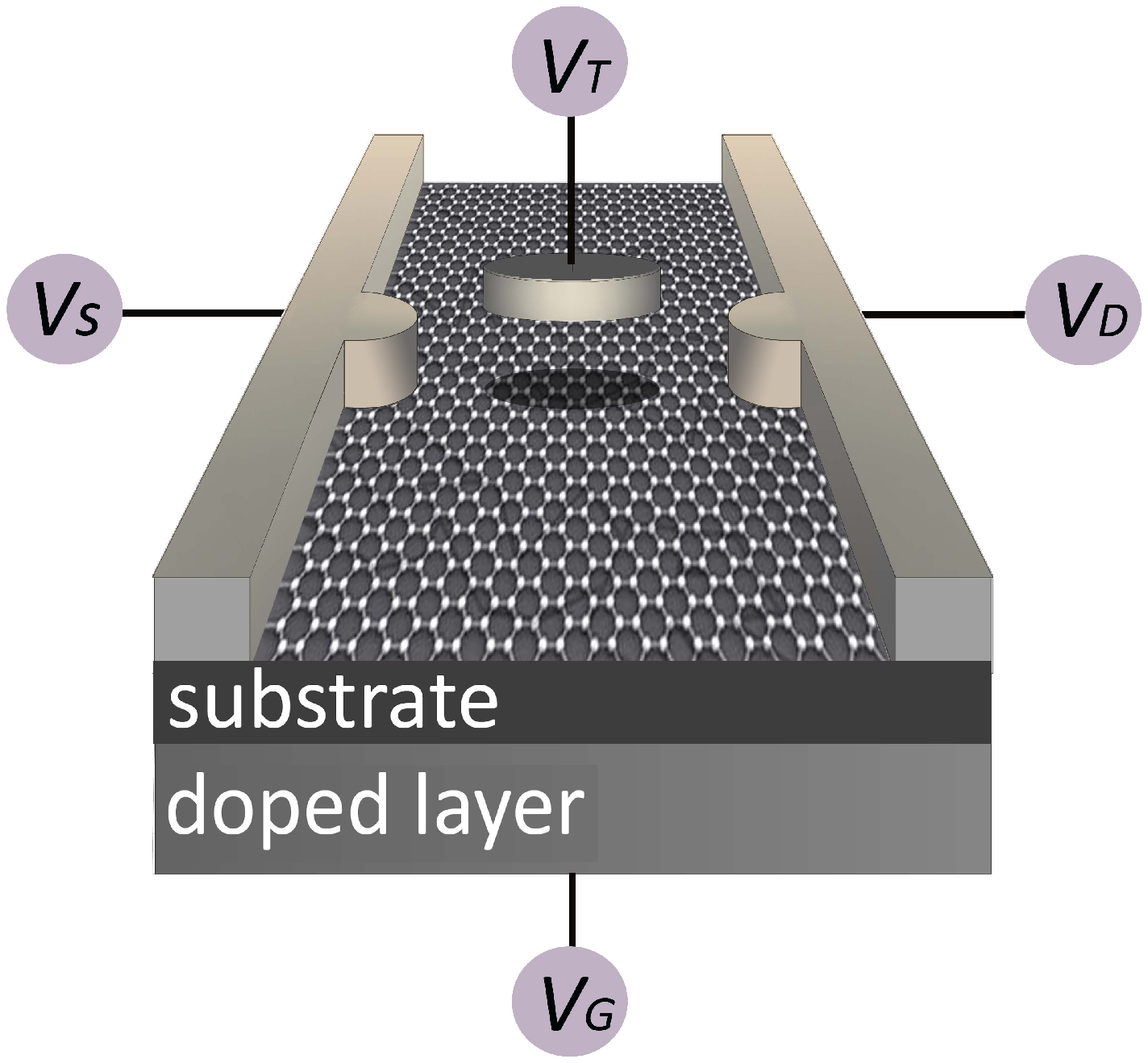,width=1.65in,,angle=0}}
\put(4.6,3.2){\epsfig{file=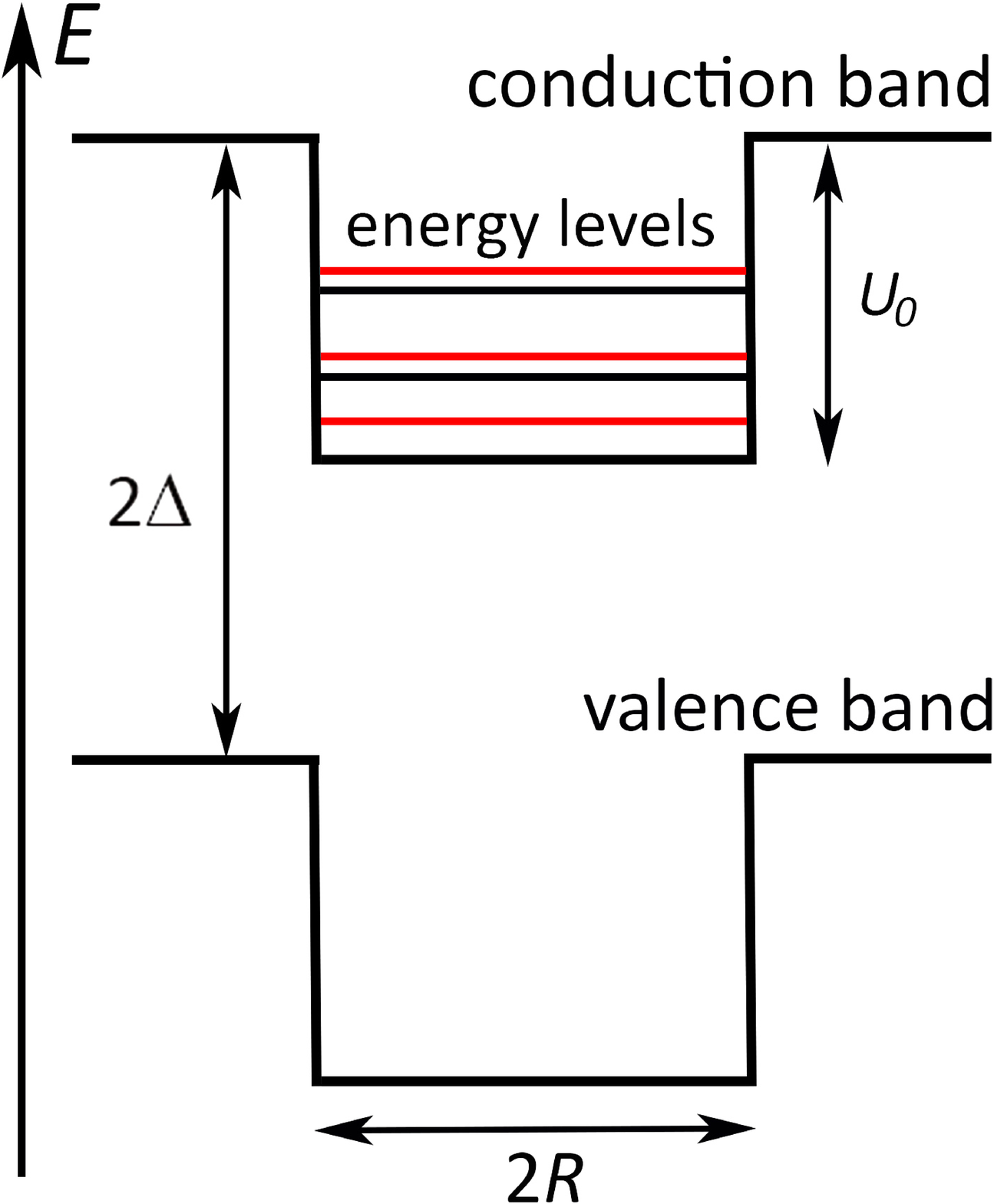,width=1.3in,,angle=0}}
\put(0.6,-0.4){\epsfig{file=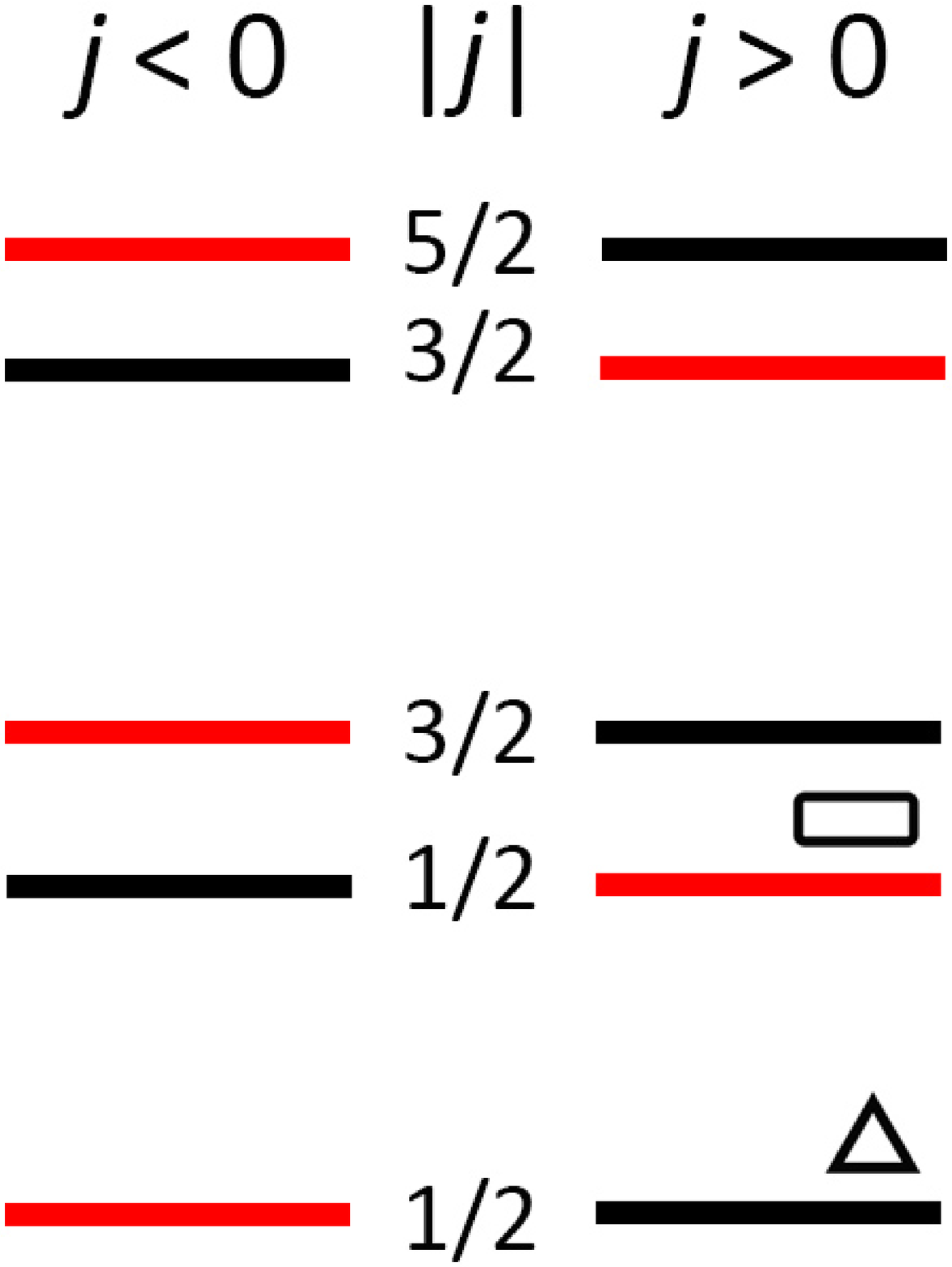,width=1.1in,,angle=0}}
\put(3.8,-0.3){\epsfig{file=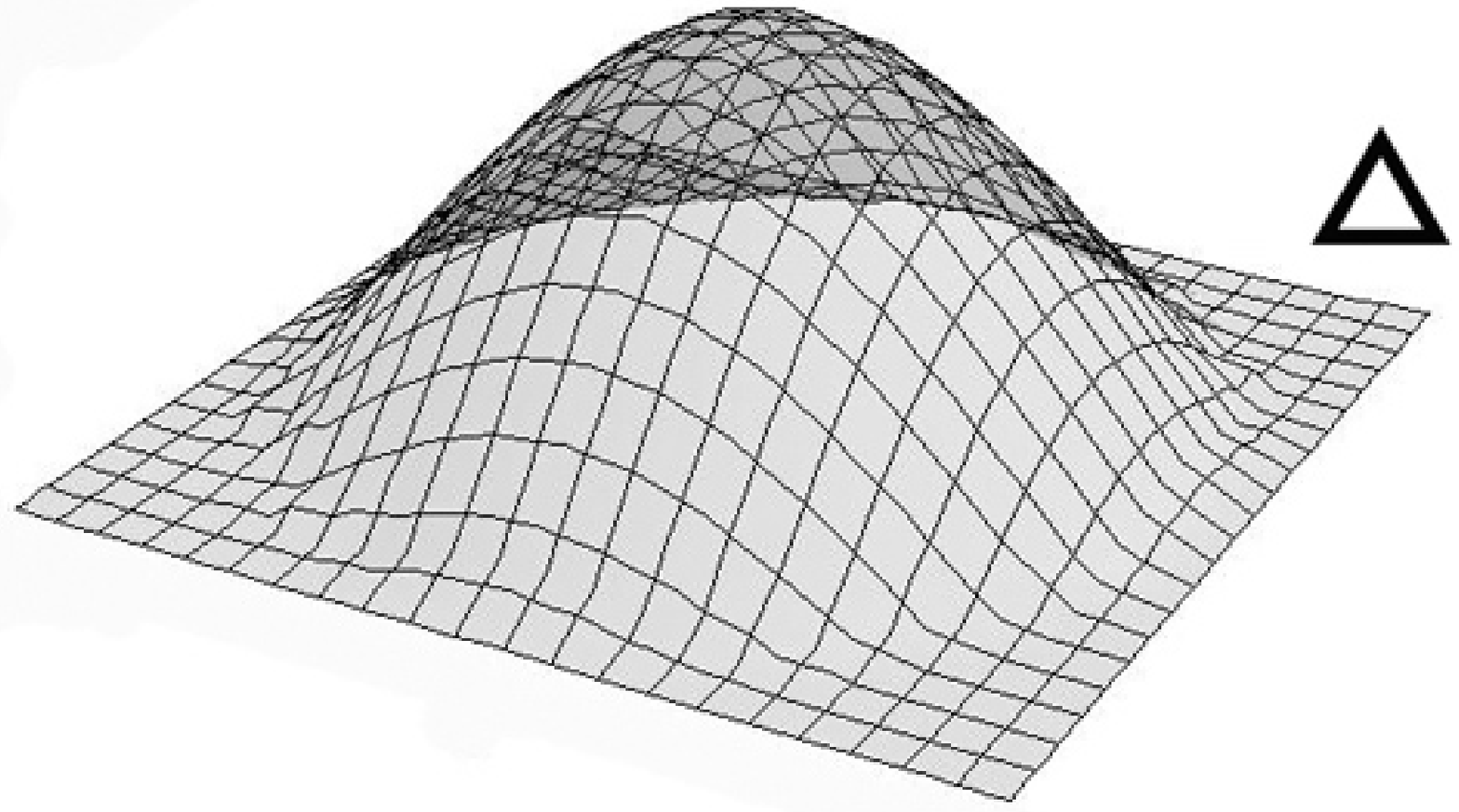,width=0.82in,,angle=0}}
\put(3.8,1.3){\epsfig{file=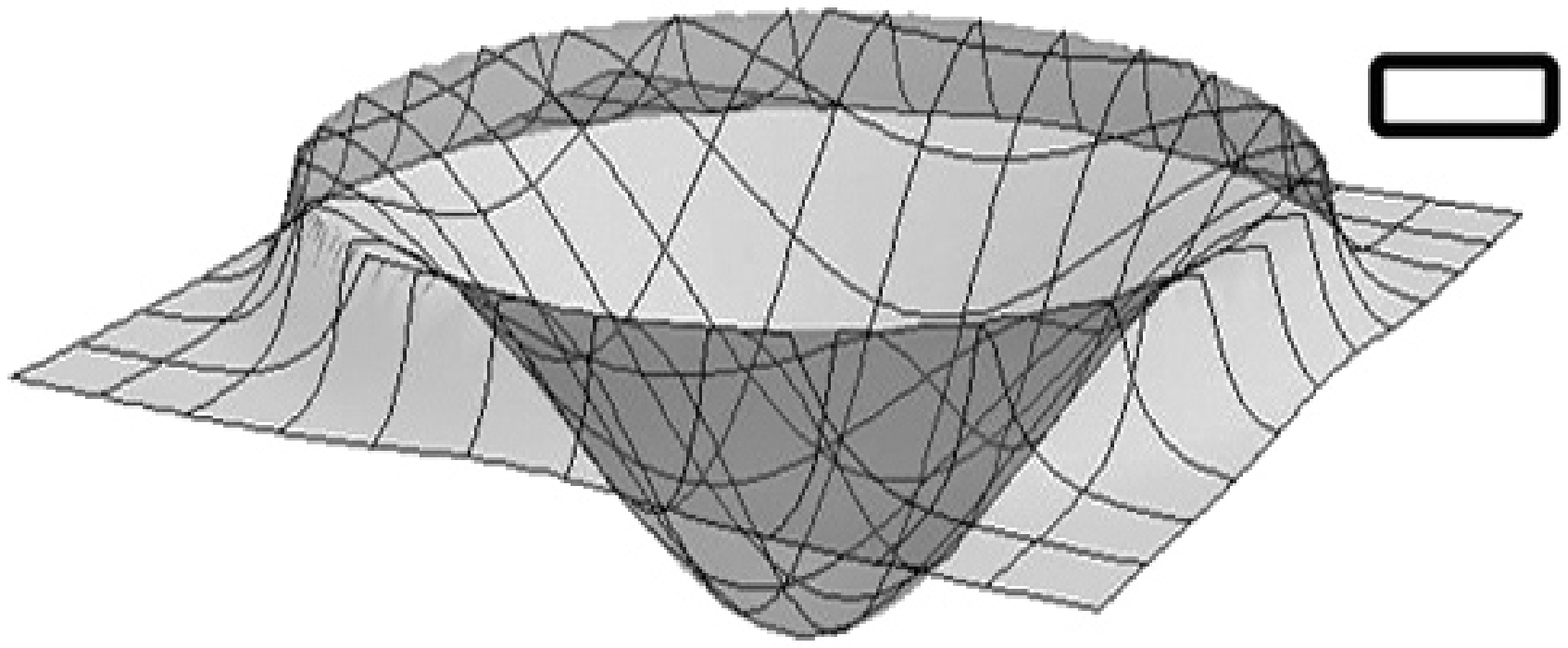,width=0.82in,,angle=0}}
\put(6.3,-0.3){\epsfig{file=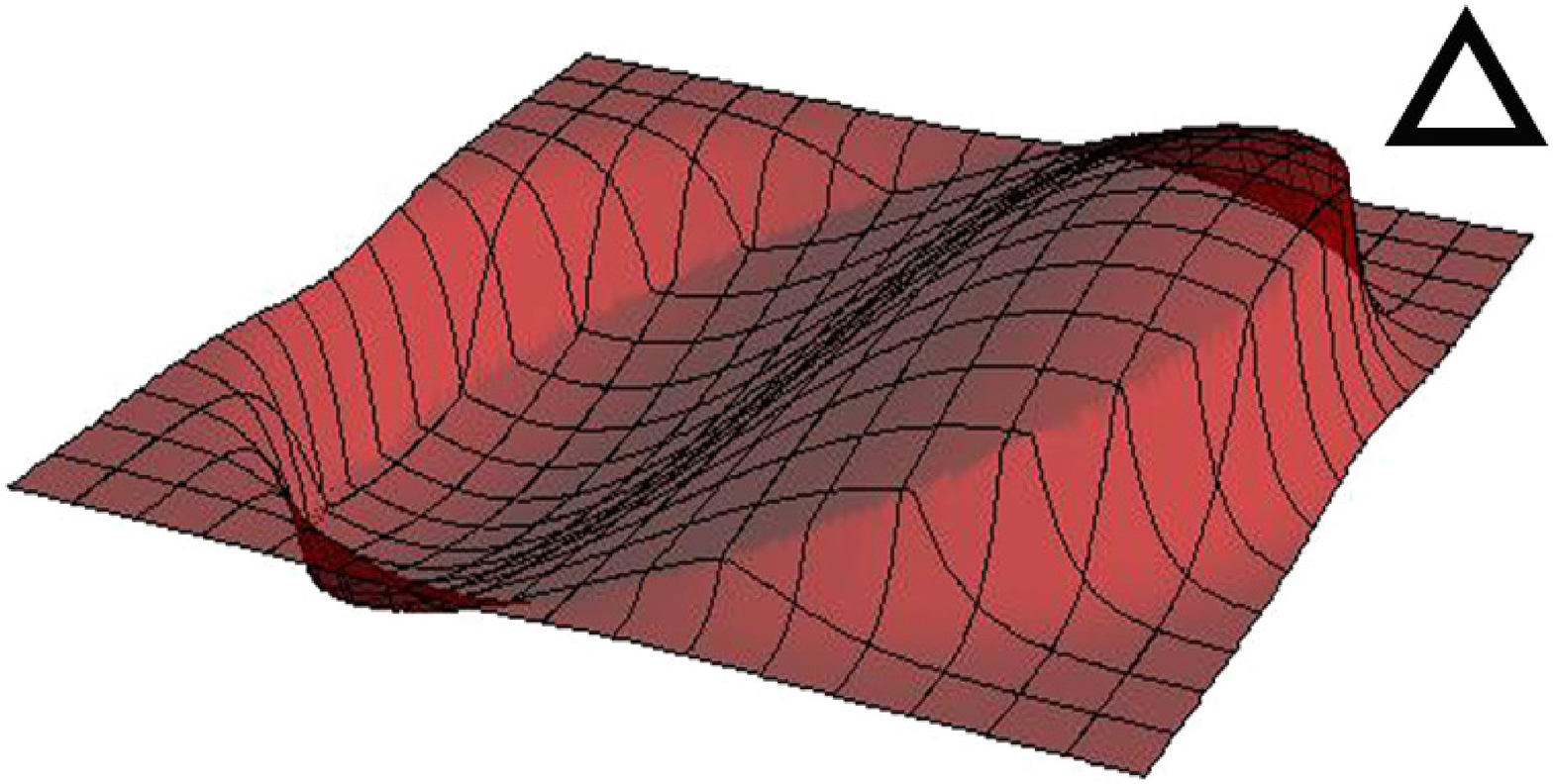,width=0.82in,,angle=0}}
\put(6.3,1.14){\epsfig{file=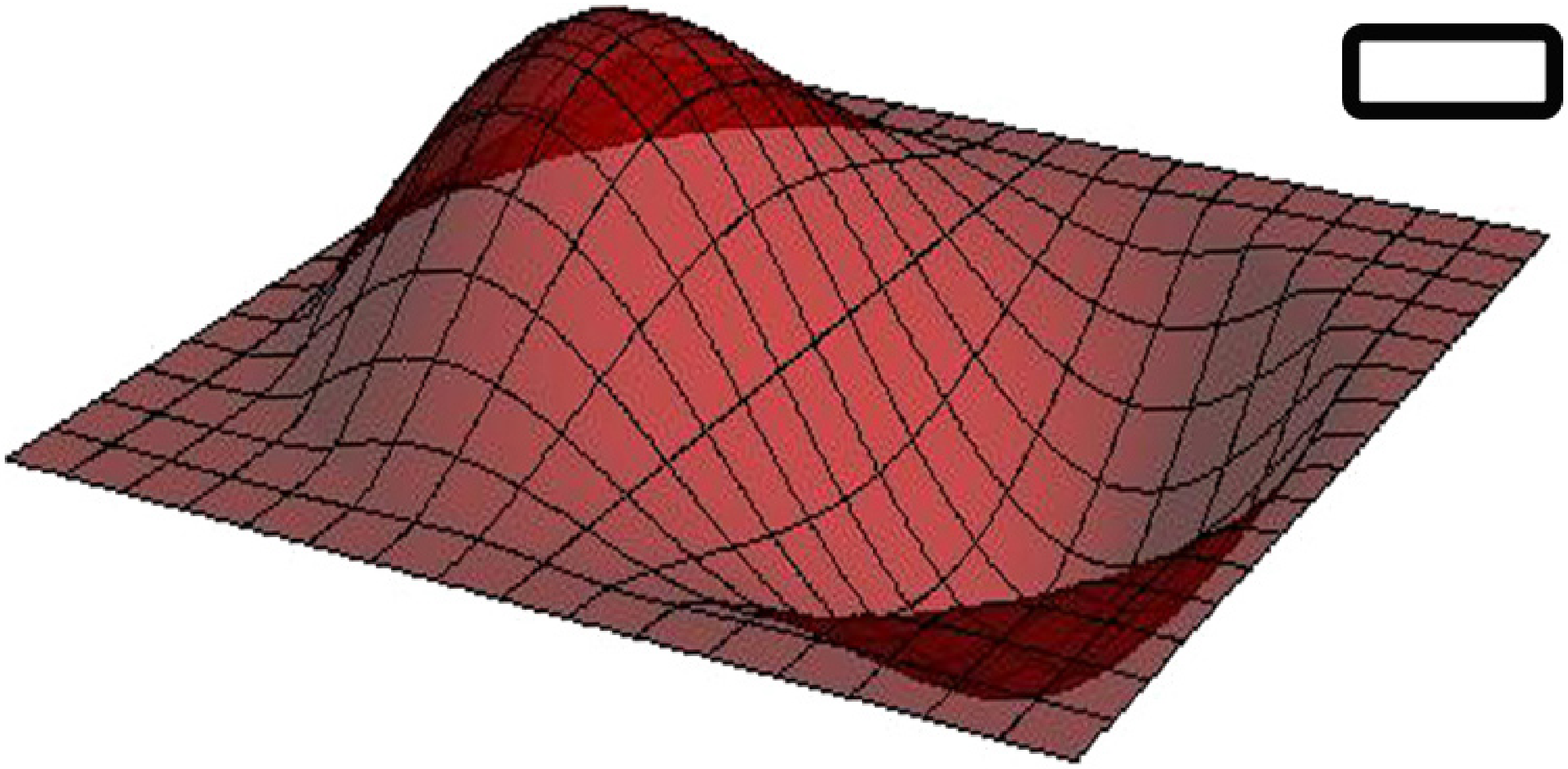,width=0.82in,,angle=0}}
\put(0.1,6.9){(a)}
\put(0.1,-0.2){(c)}
\put(8.1,-0.2){(d)}
\put(8.1,7.0){(b)}
\put(4.7,2.3){A}
\put(7.3,2.3){B}
\put(3.8,1.3){$\times 10$}
\put(6.5,-0.3){$\times 100$}
\end{picture}
\caption{(color online)
Graphene QD defined by electrostatic gates.
(a) Proposed setup. 
The top gate (V$_T$) defines the dot
while source (V$_S$), drain (V$_D$), and back 
(V$_G$) gates allow for Coulomb blockade spectroscopy. 
(b) Radial QD confinement potential. 
The interaction between graphene and substrate 
opens a mass gap $2\Delta$ in the QD energy spectrum.
(c) Lowest noninteracting 
QD energy levels in the conduction band.
Black (red [gray]) lines label states in K (K$'$) valley. 
(d) Real part of sublattice-resolved envelopes
whose energies are labeled by the
square and triangle symbols 
in panel c.    
}
\label{f:QD}
\end{figure}

However, electrons in bulk graphene allegedly behave as noninteracting 
particles, except for subtle effects due to velocity 
renormalization,\cite{Elias2011,Chae2012,Siegel2013}
coupling with phonons / plasmons,\cite{Bostwick2010} 
and a hypothetical excitonic gap.\cite{Khveshchenko2001,Drut2009,Rontani2013}
The key to this paradox is that the density parameter $r_s$,
which quantifies the impact of electron correlations,\cite{Ashcroft1976}
does not depend on the electron density $n$ but coincides 
with $\alpha$.\cite{absence_WC:article}
In contrast, $r_s\sim n^{-1/2}$ of the conventional
two-dimensional electron gas \cite{Ando1982}  
increases as $n$ decreases
due to the massive dispersion of electrons. An electron solid
(Wigner crystal) is even predicted in 
the dilute limit,\cite{Wigner:article} as the
long-range order induced by Coulomb interaction localizes electrons in space.
Therefore, a way to disclose the many-body physics of graphene
is to make electrons massive, invalidating the above scaling argument.  
This occurs e.g. in the fractional 
quantum Hall effect \cite{Du2009,Bolotin2009}
and in bilayer graphene,\cite{McCann2013} 
which might be an excitonic insulator.\cite{Min2008,Rontani2013}

In this paper we explore theoretically the few-body physics
of a graphene QD with a mass gap. Our motivation is twofold:
On one side, electrons in semiconductor QDs 
may form Wigner molecules, 
\cite{Reimann2002,Ellenberger2006,Kalliakos2008,Singha2010} 
i.e., finite-size
precursors of the Wigner crystal, including carbon-based 
nanostructures---nanotubes---for which
the effect is dramatic.\cite{Pecker2013} On the other side,
a current trend in graphene QDs 
is to minimize the roles of disorder and edge states, 
which are extrinsic sources of localization. These next-generation 
devices include atomically 
precise nanoribbons \cite{Wang2011,Ruffieux2012} and 
bilayer QDs---possibly defined 
through 
gates.\cite{MiltonPereira2007,Allen2012,Goossens2012,Muller2013,Zarenia2013}

Here we consider a clean, circular QD
with a mass gap induced by the breaking of sublattice symmetry. 
This could be realized through
the interaction between graphene 
and substrate,\cite{Recher2009} such as  
BN \cite{Giovannetti2007,Amet2013,Hunt2013,Woods2014} 
and SiC \cite{Zhou2007}
(but the evidence for these materials is debated \cite{Dean2010,Xue2011,Decker2011,Woods2014,Vitali2008,Rotenberg2008,ZhouII2008}).
The presence of the gap allows to electrostatically define the QD
as well as to perform Coulomb blockade spectroscopy,
as sketched in Figs.~\ref{f:QD}(a)-(b).

Other authors already suggested that electrons in graphene QDs may 
crystallize. However, some of these analyses were limited to degenerate
edge states that are sensitive to interactions as well as to all kinds
of perturbation,\cite{Wunsch2008,Romanovsky2009,Guclu2009,Potasz2012} 
whereas other theories treated Coulomb interaction 
at the mean field level,\cite{Paananen2011} which
may artificially enhance localization,\cite{Reimann2002} or
considered only valley-polarized electrons,\cite{Yang2012} 
which artfully breaks time-reversal symmetry. 
Here we exactly diagonalize 
the full interacting Hamiltonian 
taking into account correlations at all
orders and the presence of inequivalent K (isospin $\tau=1$)
and K$^{\prime}$ ($\tau=-1$) Dirac cones.
Through the analysis of the energy spectrum, charge density
and pair correlation functions we
show that electrons form Wigner molecules in realistic devices,
exhibiting signatures of crystallization in Coulomb blockade spectra.

The structure of this paper is the following:
After illustrating the low-energy effective-mass Hamiltonian
as well as the exact-diagonalization method we use to
solve the few-body problem (Sec.~\ref{s:theo}),
we report our predictions for the QD addition energy (Sec.~\ref{s:add})
and one-body charge density (Sec.~\ref{s:rho}). These data,
together with spin-resolved charge densities (Sec.~\ref{s:spin}),
show evidence of Wigner localization in a broad range of device parameters.
By breaking the QD circular symmetry through angular pair correlation
functions (Sec.~\ref{s:ang}) we are able to image the formation of
Wigner molecules in space. We predict as an experimental signature 
of the Wigner molecule the quenching of its highly degenerate excitation 
energies (Sec.~\ref{s:exc}).

\section{Theoretical model}\label{s:theo}

The envelope-function QD Hamiltonian for
noninteracting electrons in the valley $\tau$ (Ref.~\onlinecite{Recher2009}) 
is
\begin{equation} \label{eq:SP_hamiltonian}  
\hat{H}_{\tau} = -i\hbar v_{F}\left( 
\hat{\tau}_1\partial/\partial x + \hat{\tau}_2
\partial/\partial y \right)  +\tau\Delta\hat{\tau}_3 
+ U\!(\rho) \hat{\tau}_0.
\end{equation}
Here $v_{F}\approx10^{6} $ m/s is the Fermi velocity, 
the $2 \times 2$ Pauli matrices $\hat{\tau}_1$,
$\hat{\tau}_2$, $\hat{\tau}_3$,
and the unit matrix $\hat{\tau}_0$ act on 
pseudospinors whose components are the A/B 
sublattice envelopes, $U\!(\rho)=U_{0}\,\Theta(\rho-R)$ 
is the circular hard-wall confinement potential of height $U_0$
plotted in Fig.~\ref{f:QD}(b), with
$R$ being the QD radius 
and $\rho=(x^2+y^2)^{1/2}$.
The potential $U$,  
modulated by the top gate shown in Fig.~\ref{f:QD}(a),
confines the electrons in the QD since
the Zeeman-like term $\Delta\hat{\tau}_3$ 
breaks sublattice inversion symmetry, 
hence inducing a gap $2\Delta$ into the 
energy spectrum [Fig.~\ref{f:QD}(b)]. 
In the following we take $\Delta=U_{0}= 0.26$ eV.

We find numerically the eigenvalues of $\hat{H}_{\tau}$ 
following the method of Ref.~\onlinecite{Recher2009}.
The QD bound states 
$\Phi(\mathbf{r})$ are pseudospinors of the form
\begin{equation} \label{eq:SP_spinor}
\Phi(\mathbf{r}) 
= e^{i (j-1/2)\varphi} 
\begin{pmatrix} 
\mathcal{R}^{A} (\rho) \\ 
 \\
e^{i \varphi}  \mathcal{R}^{B} (\rho)
\end{pmatrix} ,
\end{equation}
where $\varphi$ is the azimuthal angle,
$j=\pm 1/2$, $\pm 3/2$, $\ldots$ is the 
half-integer quantum number eigenvalue
of the total angular momentum 
$\hat{\j}_{z} = -i\hbar\partial/\partial \varphi
+\hbar \hat{\tau}_{3}/2 $,
and $\mathcal{R}^{A} (\rho)$  
[$\mathcal{R}^{B} (\rho)$]
is the radial envelope on sublattice A [B] 
(Ref.~\onlinecite{DiVincenzo1984}).
As illustrated in Fig.~\ref{f:QD}(c) for the lowest 
conduction-band states, QD orbitals
whose quantum numbers differ solely in the sign of
$\tau$ (black or red [gray] lines) 
have different energies since inversion symmetry is broken,
whereas time reversal simmetry protects  
$\varepsilon(\tau,j)=\varepsilon(-\tau,-j)$.
Overall, including the spin degree of freedom $\sigma = \uparrow$,
$\downarrow$, QD levels are four-fold degenerate.
Both radial profiles and integrated weights 
of envelopes $\mathcal{R}(\rho)$ are
generically different on the two
sublattices,
as shown in the example of Fig.~\ref{f:QD}(d). 

We consider a few excess interacting charge carriers
populating the QD conduction band.
The presence of the gap $2\Delta$ 
allows us to ignore the pathologies that plague the many-body
problem of Dirac electrons due to the unboundedness 
of the energy spectrum.\cite{Greiner1985,Hausler2009}
The interacting Hamiltonian is
\begin{eqnarray} \label{eq:MB_hamiltonian}
\hat{H} &=& \sum_{a\tau\sigma}\varepsilon_{a\tau}
\hat{c}^{\dagger}_{a\tau\sigma}\hat{c}_{a\tau\sigma}
+\dfrac{1}{2}\sum_{abcd}\sum_{\tau\tau'}\sum_{\sigma\sigma'}\nonumber\\
&\times&
\left<a\tau, b\tau'\right|v(\mathbf{r}
-\mathbf{r'})\left|c\tau', d\tau\right>
\hat{c}^{\dagger}_{a\tau\sigma}
\hat{c}^{\dagger}_{b\tau'\sigma'}\hat{c}_{c\tau'\sigma'}
\hat{c}_{d\tau\sigma},
\end{eqnarray} 
where $\hat{c}^{\dagger}_{a\tau\sigma}$ 
creates an electron of spin 
$\sigma$ in the orbital $\left|a\tau \right>$
labeled by quantum numbers $\tau$ and $a\equiv(j_a,n_a)$ 
whose energy is $\varepsilon_{a\tau}$
($n_a$ is the number of radial nodes).
Two-body interaction takes the Ohno form
$v(\mathbf{r} -\mathbf{r'}) = v_0\left[ 1+(v_0\epsilon/e^2)^2
|\mathbf{r}-\mathbf{r'}|^2\right]^{-1/2}$, where
$\epsilon$ is the background relative dielectric constant.
Since realistic values of $\epsilon$ fall in a wide range between 
$\epsilon=1.4$ and $\epsilon=44$, depending on the substrate 
\cite{Walter2011,Hwang2012} as well as on nearby gates,
here we treat $\epsilon$ as a free parameter.
At large distances $v$ approaches the Coulomb potential,
whereas its contact limit is the Hubbard-like intra-atomic interaction
$v_{0}=15$ eV for the $2p_{z}$ orbital.\cite{Ohno1964}
Matrix elements $\left<a\tau, b\tau'|v|c\tau', d\tau\right>$
are obtained from tight-binding states 
neglecting interatomic orbital overlaps 
\cite{Secchi2010} as well as
small intervalley exchange terms.\cite{Secchi2013}

The many-body states 
are superpositions of the Slater determinants obtained by
filling the lowest 68 spin-valley-orbitals with $N$ 
electrons in all possible ways (aka full configuration interaction 
\cite{Rontani_FullCI2006}). This size of the truncated single-particle
basis set was chosen after checking that the computed
many-body ground-state energy is well converged. 
In the Fock basis of Slater determinants $\hat{H}$ 
is a sparse matrix,
with blocks labeled by the total angular momentum and (iso)spin.
The maximum linear size of the matrix is 2,187,712,
which we diagonalize with the home-built parallel 
code DONRODRIGO.\cite{Rontani_FullCI2006,Kalliakos2008,Singha2010,Pecker2013}
This provides highly accurate energies and wave functions of both ground
and excited states, in contrast to other high-level methods, such as
quantum Monte Carlo, addressing ground state properties only.

\begin{figure}
\setlength{\unitlength}{1 cm}
\begin{picture}(8.5,6.5)
\put(-0.2,-0.2){\epsfig{file= 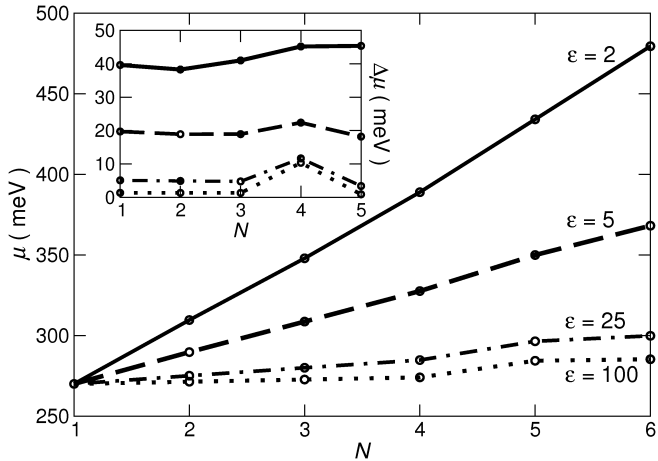,width=3.4in,,angle=0}}
\end{picture}
\caption{Coulomb blockade linear spectroscopy.
Chemical potential $\mu(N)$ vs electron number $N$ for different
background dielectric constants
$\epsilon$,
with radius $R=250$ \AA.
Inset: Charging energy $\Delta\mu(N)$ vs $N$.
Lines are guides to the eye.
$\Delta\mu$ may be measured
as electrons are added to the quantum dot one by one
tuning the backgate shown in Fig.~\ref{f:QD}(a).
}
\label{f:addition}
\end{figure}

\section{Coulomb blockade spectroscopy}\label{s:add}

A key quantity we obtain from the computed ground state energies
$E_0(N)$ is the
chemical potential $\mu(N)=E_0(N)-E_0(N-1)$, that is the resonating 
tunneling energy of the $N$th electron injected into the QD 
containing $N-1$ interacting particles. This may be measured
through Coulomb blockade spectroscopy,
as electrons are added to the QD one by one
tuning the backgate shown in Fig.~\ref{f:QD}(a).\cite{Kouwenhoven1997b}
In Fig.~\ref{f:addition} we artificially modulate 
the background screening $\epsilon$ 
to highlight the effect of Coulomb interaction
on the filling sequence (here $R=250$ \AA).
In the absence of interactions ($\epsilon=100$, dotted line),
$\mu(N)$ is constant except for a step when adding the fifth electron,
which corresponds to a peak in the charging energy $\Delta\mu(N)=
\mu(N+1)-\mu(N)$ (see inset). 
This finite value $\Delta\mu(N=4)\approx 10$ 
meV is the orbital energy cost required to add an electron 
to the second 
shell after the first one has been filled with four electrons.
This fourfold periodicity is generic for all fillings,
as clear from Fig.~\ref{f:QD}(c).

As the interaction strength is turned on,
the shell structure of $\mu(N)$ is progressively washed out. 
In contrast with circular QDs in
ordinary semiconductors,\cite{Tarucha1996}
the charging energy $\Delta \mu$ shown in the inset of
Fig.~\ref{f:addition} neither
exhibits half-shell peaks linked to Hund's rule
nor decreases with $N$. The former feature, shared by
carbon-nanotube QDs,\cite{Secchi2009,Pecker2013} is due to
the spin-valley multicomponent nature of the wave
function. In fact, at the noninteracting level
the four-fold degenerate spin-valley projections
are linked to a single orbital state, hence there is no Hund's rule,
which is associated with the partial filling of a degenerate manifold
of separate orbital states. The latter feature is peculiar to
the hard-wall confinement potential, as in the case
of ordinary semiconductors the potential is soft
so the dot size $L$ increases with $N$
whereas the charging energy $\Delta \mu =e^2 / C$ decreases with $N$
($C\sim L$ is the QD capacitance).

For realistic values of $\epsilon$ 
the Coulomb energy overwhelms the kinetic energy,
making $\mu$ increase almost linearly with $N$
(dashed and solid lines 
in Fig.~\ref{f:addition} for $\epsilon=5$ and 2,
respectively). 

\begin{figure}
\setlength{\unitlength}{1 cm}
\begin{picture}(8.5,6.5)
\put(-0.2,-0.2){\epsfig{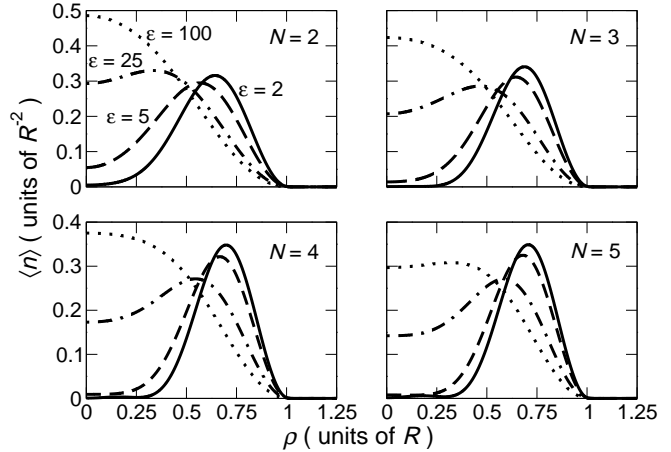}}
\end{picture}
\caption{Emergence of radial correlations in the wave function.
One-body density $\left<n(\mathbf{r})\right>$
vs radial coordinate $\rho$ for different values of
dielectric constant
$\epsilon$ and electron number
$N$, with radius
$R=1250$ \AA. Realistically screened mutual interactions
push electrons against the QD potential wall.
}
\label{f:onebody}
\end{figure}

\section{Emergence of radial correlations}\label{s:rho}

To clarify how interactions affect the 
wave function we compute the---circularly symmetric---one-body density
\begin{equation} \label{eq:one_body_density}
\left< n(\mathbf{r})\right> = \frac{1}{N}\sum_{i=1}^N  
\left< \delta(\mathbf{r}-\mathbf{r}_i)\right>,
\end{equation}
where $\left<\ldots\right>$ is the quantum statical average
for vanishing temperature.
In practice, we average $\left< n(\mathbf{r})\right>$
over the ground-state multiplet, whose large  
angular-momentum degeneracy is protected
by symmetry against the effect of interaction.
After the averaging $\left< n(\mathbf{r})\right>$
is the same on both sublattices, 
unspecified in the following.

Figure \ref{f:onebody} shows the evolution of the
radial profile of $\left<n(\mathbf{r})\right>$
with the interaction strength. 
Whereas for large screening (dotted lines) the probability weight
is spread all over the QD, as $\epsilon$ is decreased
the central region is depleted with its weight being moved
towards the dot wall. For realistic screening (dashed and solid lines)
$\left<n\right>$  is a ring with electrons pushed against the potential wall
by Coulomb repulsion, which hints to the formation of a Wigner molecule. 
\cite{Reimann2002}
This trend is generic for different electron
numbers and dot radii,  
the larger $R$ the higher $\epsilon$ at which the ring structure
sets in (data not shown).

\begin{figure}
\setlength{\unitlength}{1 cm}
\begin{picture}(8.5,3.5)
\put(0.0,0.0){\epsfig{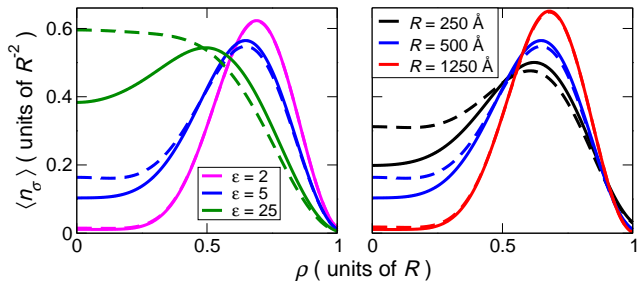}}
\end{picture}
\caption{(color online). 
Suppression of exchange interactions.
Spin-resolved density $\left< n_{\sigma}(\mathbf{r})\right>$
vs radial coordinate $\rho$ for different values of
dielectric constant $\epsilon$ (left panel,
$R=500$ \AA) and radius $R$ (right panel, $\epsilon=5$), with $N=5$
and spin projection $S_z=1/2$.
Solid and dashed lines point to
$\left<n_{\uparrow}\right>$ and $\left<n_{\downarrow}\right>$,
respectively.
Wigner localization
depletes the probability weight in the regions halfway
an electron and its neighbors and hence suppresses exchange interactions,
inducing large spin degeneracies.
}
\label{f:spinresolved}
\end{figure}

\section{Suppression of exchange interactions}\label{s:spin}

A fingerprint of Wigner localization is provided by 
the spin-resolved one-body density $\left<n_{\sigma}
(\mathbf{r})\right>$,
\begin{equation} \label{eq:one_body_density_sr}
\left< n_{\sigma}(\mathbf{r})\right> = \frac{1}{N_{\sigma}}\sum_{i=1}^N  
\left< \delta_{\sigma\sigma_i}\delta(\mathbf{r}-\mathbf{r}_i)\right>_{S_z}.
\end{equation}
Here $N_{\sigma}$ is the number of
electrons with spin $\sigma$ so $\left<n_{\sigma}
(\mathbf{r})\right>$ is normalized to one, and
$\left<\ldots\right>_{S_z}$ is the average taken over the 
manifold of states
with fixed total spin projection $S_z=(N_{\uparrow}-N_{\downarrow})/2$.
For odd electron numbers
$\left<n_{\uparrow}(\mathbf{r})\right>$
and $\left<n_{\downarrow}(\mathbf{r})\right>$
generically differ, as illustrated in Fig.~\ref{f:spinresolved}
for five electrons and $S_z=1/2$.
However, as the interaction strength is increased
by either suppressing screening (left panel) or increasing the dot size
(right panel), the radial profiles 
of $\left<n_{\uparrow}\right>$ (solid lines)
and $\left<n_{\downarrow}\right>$ (dashed lines) 
tend to overlap and form the same 
probability density ring.\cite{Kalliakos2008}
The rationale is that Coulomb forces localize electrons in space,
depleting the probability weight in the regions halfway 
an electron and its neighbors.
Therefore, 
exchange interactions between pairs of electrons are suppressed,
making spin degrees of freedom redundant. 

\section{Emergence of angular correlations}\label{s:ang}

To detect whether angular correlations are enforced by
interactions we break 
the circular symmetry of the one-body density 
introducing the pair correlation function $P(\mathbf{r}, \mathbf{r}_{0})$,
i.e., the conditional probability of finding an electron at 
$\mathbf{r}$ provided another electron is located at the fixed
position $\mathbf{r}_{0}$ displaced from the origin,
\begin{eqnarray} \label{eq:two_body_density}
P(\mathbf{r}, \mathbf{r}_0) 
&=&
\dfrac{1}{N(N-1)} 
\sum_{\sigma_1,\sigma_2,\ldots,\sigma_N}
\int\!\! d\mathbf{r}_3\, d\mathbf{r}_4\ldots d\mathbf{r}_N\nonumber\\
&&
\left|\psi(\mathbf{r},\sigma_1;\mathbf{r}_0,\sigma_2;
\mathbf{r}_3,\sigma_3;\ldots;\mathbf{r}_N,\sigma_N)\right|^2.
\end{eqnarray}
For the sake of simplicity, here
we take the quantum average 
over a selected pure quantum state $\psi$ belonging
to the ground-state multiplet and show the sublattice component
with the largest weight.

\begin{figure}
\setlength{\unitlength}{1 cm}
\begin{picture}(8.5,10.8)
\put(0.2, -0.2){\epsfig{file=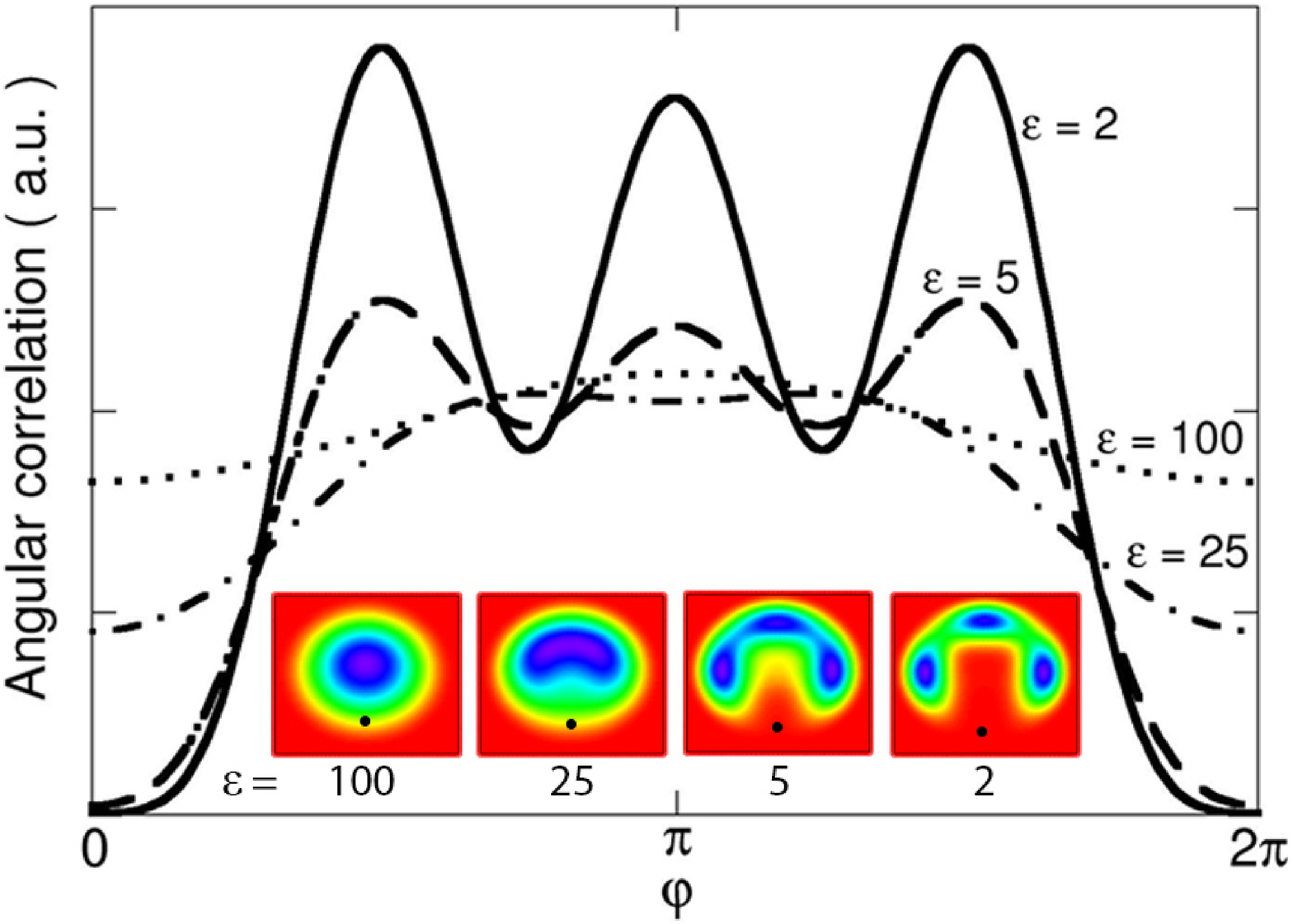,width=3.1in,,angle=0}}
\put(1.3,5.5){\epsfig{file=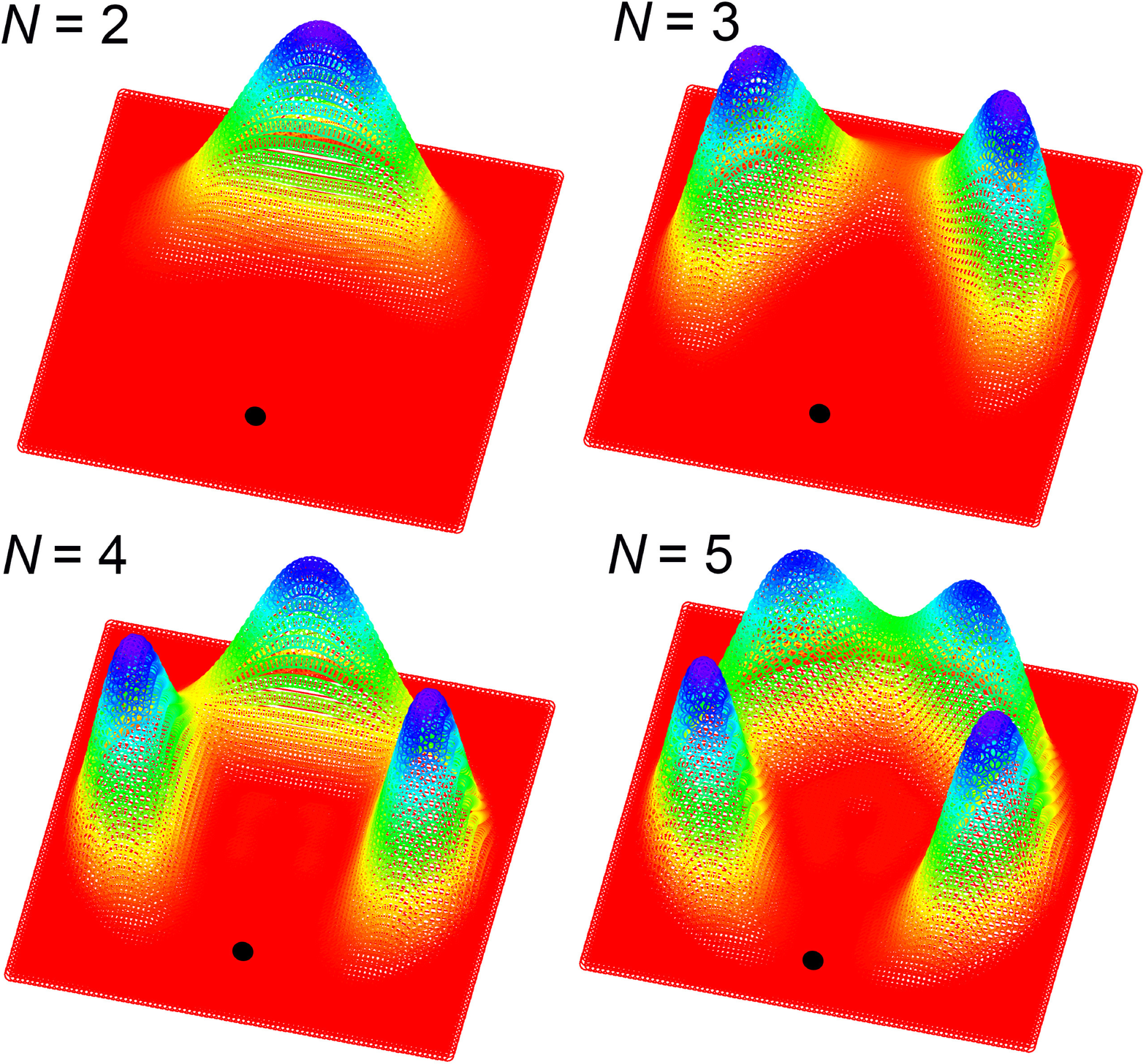,width=2.4in,,angle=0}}
\put(7.0,10.8){(a)}
\put(7.0,4.8){(b)}
\end{picture}
\caption{(color online). 
Polygonal Wigner molecules. (a) 
Three-dimensional
contour plots of pair correlation functions $P(\mathbf{r}, \mathbf{r}_0)$
for $\varepsilon=2$ and $R=2250$ \AA. Black dots point to the locations
$\mathbf{r}_0$
of fixed electrons.
(b) 
Pair correlation function
$P(\mathbf{r}, \mathbf{r}_0)$ vs angle $\varphi$
with $\left|\mathbf{r}\right| =
\left|\mathbf{r}_0\right|$ for four electrons and different values of
dielectric constant $\epsilon$,
with $R = 500$ \AA.
Inset: corresponding contour plots of $P(\mathbf{r}, \mathbf{r}_0)$
in the $xy$ plane. Increasing the interaction strength leads to
the formation of the correlation hole as well as the development
of angular correlations, which enforce a square Wigner molecule.
}
\label{f:twobody}
\end{figure}

The insets of Fig.~\ref{f:twobody}(b) show how
the contour plots of
$P(\mathbf{r}, \mathbf{r}_0)$ for four electrons
evolve in the $xy$ plane as screening is suppressed.
The black dots highlight the positions 
$\mathbf{r}_0$ of the fixed electron, located at the maximum of 
of the one-body density with arbitrary angle.   
As the interaction strength increases [panels from left ($\epsilon=100$) to
right ($\epsilon=2$)], we see---beyond the onset of the correlation 
hole around the fixed particle---a strong rearrangement of
the probability weight:  
a non-trivial structure emerges
made of three peaks located at the vertices of a square
whose last vertex is placed at $\mathbf{r}_0$. Overall, the three peaks
plus the fixed electron realize a square Wigner molecule, which rotates
together with $\mathbf{r}_0$. 

Cutting the contour plots
of $P(\mathbf{r}, \mathbf{r}_0)$ along a ring of radius
$\left|\mathbf{r}_0\right|$
allows us to appreciate the role of interactions
in driving spatial order and localization,
as we show in Fig.~\ref{f:twobody}(b).
For weak correlations (dotted line) 
$P$
vs $\varphi$
is featureless, exhibiting a minor depression close to
$\varphi=0$, $2\pi$, which realizes the exchange hole around
the fixed electron position. Increasing the interaction (up to
$\epsilon=2$, solid line) the three peaks of the square Wigner molecule
emerge together with a deep correlation hole around $\mathbf{r}_0$, 
the peak-to-valley ratio increasing with decreasing $\epsilon$.

Figure \ref{f:twobody}(a) shows the generic behavior
of $N$ electrons in
the strongly correlated limit, 
here enforced with $\epsilon=2$ and $R=2250$ \AA.
The electrons 
realize Wigner molecules
whose forms are regular polygons with $N$ vertices,
as illustrated by
the three-dimensional plots of $P(\mathbf{r}, \mathbf{r}_0)$
for the dimer ($N=2$), the triangle ($N=3$), 
the square ($N=4$), and the pentagon ($N=5$).

\begin{figure}
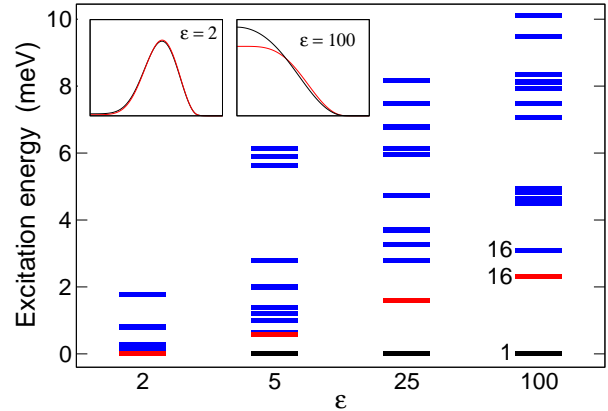

\setlength{\unitlength}{1 cm}
\begin{picture}(8.5,5.5)
\put(0.3,-0.1){\epsfig{file= ./fig6a.eps,width=3.1in,,angle=0}}
\put(1.3,3.8){\epsfig{file=./fig6b.eps,width=0.7in,,angle=0}}
\put(3.25,3.8){\epsfig{file=./fig6c.eps,width=0.7in,,angle=0}}
\end{picture}
\caption{(color online)
Excitation spectrum of a Wigner molecule.
Low-lying excitation energies
vs dielectric constant $\epsilon$ for $N=4$ and $R=500$ \AA.
Numbers label degeneracies of selected multiplets.
Insets: density $n(\mathbf{r})$ vs radial coordinate $\rho$ averaged over
the ground state
(black curve) and the first excited multiplet (red [gray] curve).
The Wigner-molecule ground state is highly degenerate
as localized electrons may independently flip their (iso)spins.
}
\label{f:excitation}
\end{figure}

\section{Excitation spectrum}\label{s:exc}

The excitation spectrum of a Wigner molecule may be measured by
either non-linear Coulomb blockade spectroscopy 
\cite{Pecker2013}---opening the source-drain 
bias window in the setup of Fig.~\ref{f:QD}(a)---or 
inelastic light scattering.\cite{Kalliakos2008,Singha2010}
Figure \ref{f:excitation} shows the dependence of 
low-lying excitation energies on the interaction strength
for four electrons. For weak interactions
($\epsilon=100$), the spectrum reminds us of the single-particle ladder of
levels of Fig.~\ref{f:QD}(c), as to excite the ground state one
moves an electron from the lowest completely filled shell to 
higher orbital states. Whereas in this specific case the ground state 
is non degenerate,
the excited multiplets exhibit large degeneracies 
(labeled by
numbers) linked to different (iso)spin orientations.
For stronger interactions, the lowest excitation
energies are strongly quenched as the system turns into a square Wigner
molecule. Comparing the one-body 
density $n(\mathbf{r})$ of the ground-state (black curves in the insets)
with $n(\mathbf{r})$ 
averaged over the lowest excited multiplet
(red [gray] curves), we see that the two curves overlap for strong interactions
(left inset, $\epsilon=2$). 
In fact, in the limit of perfect localization 
the Wigner-molecule ground state exhibits a huge
degeneracy since localized
electrons may independently flip their (iso)spins, 
as exchange interactions are completely suppressed.
Therefore, the energy spectrum of the Wigner molecule is
a ladder of highly-degeratate rotovibrational quanta. 
\cite{Reimann2002,Kalliakos2008,Singha2010}

\section{Conclusions}

In conclusion, 
electrons in a disorder-free graphene quantum dot
with a mass gap form Wigner molecules for a broad range 
of device parameters.
The signatures of Wigner localization may be traced in
Coulomb blockade and other electron spectroscopies.
We expect our findings to be generic to clean carbon-based nanostructures 
exhibiting a mass gap, including atomically precise 
ribbons and bilayer-graphene quantum dots.

\begin{acknowledgments}
We thank Andrea Secchi, Elisa Molinari, Deborah Prezzi, Marco Polini, 
Andrea Candini, Andrea Ferretti, Vittorio Pellegrini, Stefano Corni,
Stefan Heun,
and Pino D'Amico for stimulating discussions.
This work is supported by MIUR-PRIN2012 MEMO, EU-FP7 Marie Curie initial 
training network INDEX, MIUR ABNANOTECH,
CINECA-ISCRA grants IscrC\_TUN1DFEW, IscrC\_TRAP-DIP, and IscrC\_PAIR-1D.
\end{acknowledgments}

%

\end{document}